\documentclass[12pt, a4paper]{article}

    \usepackage{graphicx}
    \usepackage{latexsym}
    \usepackage{amsmath}
    \usepackage{amssymb}
    \usepackage{epsfig}
    \usepackage{array}

\begin{document}

    \title{Wolfgang Pauli and Modern Physics
    \footnote{Invited talk at the Workshop ``The Nature of Gravity'', at the International Space Science Institute (ISSI),
    6 - 10 October 2008, Bern, Switzerland.}}

    \author{Norbert Straumann\\
        Institute for Theoretical Physics University of Zurich,\\
        CH--8057 Zurich, Switzerland}

    \maketitle

    \begin{abstract}
    In this written version of a pre-dinner-speech at the workshop ``The Nature of Gravity'' at ISSI I illustrate Pauli's science primarily with material that has not formally been published by him, but was communicated in detailed letters to eminent colleagues and friends.
    \end{abstract}

    \section{Introduction}

    Wolfgang Pauli was one of the most influential figures in twentieth-century science. In
    the foreword of the memorial volume to Pauli, edited by Markus Fierz and Victor Weisskopf \cite{FW}
    -- two former assistants of Pauli -- Niels Bohr wrote about Pauli: ``At the same time
    as the anecdotes around his personality grew into a veritable legend, he more and more
    became the very conscience of the community of theoretical physicists.'' There are few fields
    of physics on which Pauli's ideas have not left a significant imprint. From Pauli's enormous
    correspondence, edited by Karl von Meyenn \cite{Mey}, and his studies in historical, epistemological
    and psychological questions, it becomes obvious that his searching mind embraced all aspects
    of human endeavor.

I knew Pauli only as a student. Beside attending his main courses, I saw him in action in the joint Theoretical Physics Seminar of ETH and the University, and in our general Physics Colloquium. Although it was a bit too early for me, I also visited some specialized lectures. In addition, I vividly remember a few public talks, like the famous one ``On the earlier and more recent History of the Neutrino'' that was given by Pauli immediately after the discovery of parity violation \cite{P1}. Therefore, I can only talk about Pauli's science. Pauli was obviously a difficult and complex personality. Markus Fierz, who knew and understood him particularly well, once said in a talk: {\it``Whoever knew him also felt that in this man the opposites of heavenly light and archaic darkness were having a tremendous impact.''} And in a letter Markus Fierz wrote to me that a true biography would have to be written by a physicist with poetic gifts. In certain circles there is now a lot of interest in Pauli's special relationship with the psychiatrist Carl Gustav Jung. Only few physicist colleagues knew about this and corresponded with him on Jungian ideas and psychology in general. Pauli attached great importance to the analysis of his dreams and wrote them down in great number. Hundreds of pages of Pauli's notes are not yet published. There are scholars who are convinced that Pauli's thoughts about psychology are important. Others regard all this as mystical mumbo-jumbo. At any rate, Jung successfully helped Pauli to overcome his life crisis after his first marriage had broken up in 1930.

It would be hopeless and pointless to give an overview of Pauli's most important scientific contributions, especially since Charles Enz, the last of a prestigious chain of Pauli assistants, has published a very complete scientific biography \cite{Enz}. Much of what Pauli has achieved has become an integral part of physics\footnote{For a review, see \cite{NS2}.}. You are all aware that he was one of the founders of quantum electrodynamics and quantum field theory in general, and he is, of course, the father of the neutrino. After some biographical remarks, and a sketch of his early work, I will select some important material that appeared only in letters\footnote{Hopefully, Pauli's Scientific Correspondence, admirably edited by Karl von Meyenn, will one day be translated -- at least in part -- into English. This is a source of wonderful insights.}.

\section{A brief Biography}

    Let me begin with a few biographical remarks. Pauli was born in 1900, the year of Planck's
    great discovery. During the high school years Wolfgang developed into an infant prodigy
    familiar with the mathematics and physics of his day.

    Pauli's scientific career started when he went to Munich in autumn 1918 to study theoretical
    physics with Arnold Sommerfeld, who had created a ``nursery of theoretical physics''. Just
    before he left Vienna on 22 September he had submitted his first published paper, devoted to the energy components of the gravitational field in general relativity \cite{P2}. As a 19-year-old student he then wrote two papers \cite{P3}, \cite{P4} about the recent
    brilliant unification attempt of Hermann Weyl (which can be considered in many ways as the
    origin of modern gauge theories). In one of them he computed the perihelion
    motion of Mercury and the light deflection for a field action which was then preferred by
    Weyl.  From these first papers it becomes obvious that Pauli mastered the new field completely.
    Hermann Weyl was astonished. Already on 10 May, 1919, he wrote to Pauli from Z\"{u}rich:
    \textit{``I am extremely pleased to be able to welcome you as a collaborator. However, it is almost
    inconceivable to me how you could possibly have succeeded at so young an age to get hold of
    all the means of knowledge and to acquire the liberty of thought that is needed to assimilate
    the theory of relativity.''}

    Sommerfeld immediately recognized the extraordinary talent of
    Pauli and asked him to write a chapter on relativity in {\it Encyklop\"{a}die der
    mathemati\-schen Wissenschaften}. Pauli was
    in his third term when he began to write this article. Within less than one year he finished
    this demanding job, beside his other studies at the university. With this article \cite{P5}, \cite{P6} of 237 pages
    and almost 400 digested references Pauli established himself as a scientist of rare depth and
    surpassing synthetic and critical abilities. Einstein's reaction was very positive:
    \textit{``One wonders what to admire most, the psychological understanding
    for the development of ideas, the sureness of mathematical deduction, the profound physical
    insight, the capacity for lucid, systematic presentation, the knowledge of the literature,
    the complete treatment of the subject matter or the sureness of critical appraisal.''}

    Pauli studied at the University of Munich for six semesters. At the time when his Encyclopedia
    article appeared, he obtained his doctorate with a dissertation on the hydrogen
    molecule ion $H_2^+$ in the old Bohr-Sommerfeld theory. In it the limitations of the old
    quantum theory showed up. About the faculties of the young Pauli Lise Meitner wrote to Pauli's widow Franca on 22 June 1959: \textit{``I often thought of and also have told it that in the fall of 1921 I have met Sommerfeld in Lund, and that he told me he had such a gifted student that the latter could not learn anything any more from him, but because of the university laws valid in Germany he had to sit through (\textit{absitzen}) 6 semesters in order to make his doctorate. Therefore he, Sommerfeld had set his student on an encyclopedia article (...).''} (From \cite{Enz} p. 25.)

    In the winter semester of 1921/22 Pauli was Max Born's assistant in G\"{o}ttingen. During
    this time the two collaborated on the systematic application of astronomical perturbation
    theory to atomic physics. Already on 29 November, 1921, Born wrote to Einstein: \textit{``Little
    Pauli is very stimulating: I will never have again such a good assistant.''} Well, Pauli's
    successor was Werner Heisenberg.

    \section{Discovery of the Exclusion Principle}

    Pauli's next stages were in Hamburg and Copenhagen. His work during these
    crucial years culminated with the proposal of his exclusion principle in December 1924.
    This was Pauli's most important contribution to physics, for which he received a belated
    Nobel Prize in 1945.

    The discovery story begins in fall 1922 in Copenhagen when Pauli began to concentrate his
    efforts on the problem of the anomalous Zeeman effect. He later recalled: \textit{`A colleague who met
    me strolling rather aimlessly in the beautiful streets of Copenhagen said to me in a friendly
    manner, ``You look very unhappy''; whereupon I answered fiercely, ``How can one look happy
    when he is thinking about the anomalous Zeeman effect?'' '.}

    In a Princeton address in 1946 \cite{P7}, Pauli tells us how he felt about the anomalous
    Zeeman effect in his early days:
    \begin{quote}

   \textit{ ``The anomalous type of splitting was on the one hand especially fruitful because it
    exhibited beautiful and simple laws, but on the other hand it was hardly
    understandable, since very general assumptions concerning the electron, using classical
    theory as well as quantum theory, always led to a simple triplet. A closer
    investigation of this problem left me with the feeling that it was even more
    unapproachable (...). I could not find a satisfactory solution at that time, but
    succeeded, however, in generalizing Land\'e's analysis for the simpler case (in many
    respects) of very strong magnetic fields. This early work was of decisive importance
    for the finding of the exclusion principle.''}

    \end{quote}

This is not the place to even only sketch how Pauli arrived at his exclusion principle\footnote{For a detailed description, see, e.g., \cite{NS1}, and references therein.}. At the time -- before the advent of the new quantum mechanics -- it was not at all on the horizon, because of two basic difficulties: (1) There were no general rules to translate a classical mechanical model into a coherent quantum theory, and (2) the spin
degree of freedom was unknown. It is very impressive indeed how Pauli arrived at his principle
on the basis of the fragile Bohr-Sommerfeld theory and the known spectroscopic
material.

    Initially Pauli was not sure to what extent his exclusion principle would hold good. In a
    letter to Bohr of 12 December 1924 Pauli writes \textit{``The conception, from which I start,
    is certainly nonsense. (...) However, I believe that what I am doing here is no greater
    nonsense than the hitherto existing interpretation of the complex structure. My nonsense is
    conjugate to the hitherto customary one.''} The exclusion principle was not immediately accepted,
    although it explained many facts of atomic physics. A few days after the letter to Bohr,
    Heisenberg wrote to Pauli on a postcard: \textit{``Today I have read your new work, and it is
    certain that I am the one who}  rejoices most \textit{about it, not only because you push
    the swindle to an unimagined, giddy height (by introducing}  individual \textit{electrons
    with 4 degrees of freedom) and thereby have broken all hitherto existing records of
    which you have insulted me. (...).''.}

    At the end of his final paper \cite{P8} on the way to the exclusion principle, Pauli expresses the hope that a deeper understanding of quantum mechanics might enable us to derive the exclusion principle from more fundamental hypothesis. To some extent this hope was fulfilled in the framework of quantum field theory. Pauli's much later paper from 1940 \cite{P9} on the spin-statistics connection ends with:

    \begin{quote}
    \textit{``In conclusion we wish to state, that according to our opinion the connection between spin and statistics is one of the post important applications of the special theory of relativity.''}
    \end{quote}

    For the letters of Pauli on the exclusion principle, and the reactions of his influential
    colleagues, I refer to Vol. I of the \textit{Pauli Correspondence}, edited by Karl von Meyenn
    \cite{Mey}. Some passages are translated into English in the scientific biography by
    Charles Enz \cite{Enz}.

    \paragraph{Some side remarks.}

    Let me end this brief account with some remarks about Pauli's fruitful Hamburg time. I begin with recollections of Otto Stern from a recorded interview with Res Jost -- one of my most important teachers and a later close colleague -- in Zurich that took place on December 2, 1961. During his Hamburg time, Pauli had very close collaboration with Stern, who in 1922 had become professor for physical chemistry at the University of Hamburg. On Pauli Stern said: \textit{``But, of course, it was very nice with Pauli for, although he was thus highly learned, one could all the same really discuss physics with him. And ... you know, he was not allowed to enter our laboratory, because of the Pauli effect. Don't you know the famous Pauli effect? Jost: I know it all right, but I didn't know that this led to such consequences. Stern: Yet, now, as I said, we always went eating together, he always fetched me. But he did not enter, instead he only knocked, and I then came to the door and said I'm coming. Oh yes, we were very superstitious at the time. Jost: Did something ever happen? Stern: Alas, many things did happen. The number of Pauli effects, the guaranteed (verb\"{u}rgten) Pauli effects, is enormously large.''} (Translation from \cite{Enz}, p.149.)

    In his obituary for Stern Rabi wrote: \textit{``Some of Pauli's great theoretical contributions came from Stern's suggestions, or rather questions; for example, the theory of magnetism of free electrons in metals.''} From Charly Enz and Armin Thellung -- Pauli's last two assistants -- I have learned that Pauli has also discussed the question of zero point energies extensively with Stern during his Hamburg time, before the advent of the new quantum mechanics. The following remarks may be of some interest since they are related to things discussed at this conference.

As background I recall that Planck had introduced the zero-point energy with somewhat strange arguments in
1911. The physical role of the zero-point energy was much discussed
in the early years of quantum theory. There was, for instance, a
paper by Einstein and Stern in 1913 (\cite{Ein}, Vol. 4, Doc.
11; see also the Editorial Note, p. 270-) that aroused widespread
interest. In this two arguments in favor of the zero-point energy
were given. The first had to do with the specific heat of rotating
(diatomic) molecules. The authors developed an approximate theory of
the energy of rotating molecules and came to the conclusion that the
resulting specific heat agreed much better with recent experimental
results by Arnold Eucken, if they included the zero-point energy.
The second argument was based on a new derivation of Planck's
radiation formula. In both arguments Einstein and Stern made a
number of problematic assumptions, and in fall 1913 Einstein
retracted their results. At the second Solvay Congress in late
October 1913 Einstein said that he no longer believed in the
zero-point energy, and in a letter to Ehrenfest (\cite{Ein}, Vol. 5, Doc. 481)
he wrote that the zero-point energy was ''dead as a doornail''.

In Hamburg Stern had calculated, but never published, the
vapor pressure difference between the isotopes 20 and 22 of Neon
(using Debye theory for the solid phase). He came to the conclusion that without
zero-point energy this difference would be large enough for easy
separation of the isotopes, which is not the case in reality. These
considerations penetrated into Pauli's lectures on statistical
mechanics \cite{P10} (which I attended). The theme was taken up in an
article by Enz and Thellung  \cite{ET}. This was originally written
as a birthday gift for Pauli, but because of Pauli's early death,
appeared in a memorial volume of \textit{Helv.Phys.Acta}.

From Pauli's discussions with Enz and Thellung we know that Pauli
estimated the influence of the zero-point energy of the radiation
field -- cut off at the classical electron radius -- on the radius
of the universe, and came to the conclusion that it {\it ``could not
even reach to the moon''}.

When, as a student, I heard about this, I checked Pauli's
unpublished\footnote{A trace of this is in Pauli's Handbuch article
\cite{P11} on wave mechanics in the section where he discusses the
meaning of the zero-point energy of the quantized radiation field.}
remark by doing the following little calculation (which Pauli must
have done):

In units with $\hbar=c=1$ the vacuum energy density of the radiation
field is
\[     \langle\rho\rangle_{vac} = \frac{8\pi}{(2\pi)^3}\int_0^{\omega_{max}}
    \frac{\omega}{2}\omega^2 d\omega
             =  \frac{1}{8\pi^2} \omega_{max}^4 , \]
with
\begin{displaymath}
\omega_{max} = \frac{2\pi}{\lambda_{max}} = \frac{2\pi m_e}{\alpha}.
\end{displaymath}
The corresponding radius of the Einstein universe in Eq.(2) would
then be ($M_{pl}\equiv 1/\sqrt{G}$)
\[a = \frac{\alpha^2}{(2\pi)^{\frac{2}{3}}} \frac{M_{pl}}{m_e} \frac{1}{m_e}
\sim 31 km. \] This is indeed less than the distance to the moon.
(It would be more consistent to use the curvature radius of the
static de Sitter solution; the result is the same, up to the factor
$\sqrt{3/2}$.)

Our present estimates of the vacuum energy, that possibly is responsible for an accelerated expansion of the universe, are not much better.

    \subsubsection*{Exclusion principle and the new quantum mechanics}

    On August 26, 1926, Dirac's paper containing the Fermi-Dirac distribution was
    communicated by R. Fowler to the Royal Society. This work was the basis of Fowler's
    \textit{theory of white dwarfs}. I find it remarkable that the quantum statistics of identical
    spin-1/2 particles found its first application in astrophysics. Pauli's exclusion
    principle was independently applied to \textit{statistical thermodynamics} by Fermi\footnote{According to Max Born, Pascual Jordan was actually the first who discovered
    what came to be known as the Fermi-Dirac statistics. Unfortunately, Born, who was editor
    of the \textit{Zeitschrift f\"ur Physik}, put Jordans paper into his suitcase when
    he went for half a year to America in December of 1925, and forgot about it. For further
    details on this, I refer to the interesting article \cite{Sch} by E.L. Schucking.}. In the same
    year 1926, Pauli simplified Fermi's calculations, introducing the grand canonical
    ensemble into quantum statistics. As an application he studied the behavior of a gas in
    a magnetic field (paramagnetism).

    Heisenberg and Dirac were the first who interpreted the exclusion principle in the
    context of Schr\"odinger's wave mechanics for systems of more than one particle. In
    these papers it was not yet clear how the spin had to be described in wave mechanics.
    (Heisenberg speaks of spin coordinates, but he does not say clearly what he means by
    this.) The definite formulation was soon provided by Pauli in a beautiful paper \cite{P12},
    in which he introduced his famous \textit{spin matrices} and two-component spinor wave functions.

    At this point the foundations of non-relativistic quantum mechanics had been completed
    in definite form. For a lively discussion of the role of the exclusion principle in physics
    and chemistry from this foundational period, I refer to Ehrenfest's
    opening laudation \cite{E} when Pauli received the Lorentz medal in 1931. This concluded with
    the words: \textit{``You must admit, Pauli, that if you would only partially repeal your
    prohibitions, you could relieve many of our practical worries, for example the traffic
    problem on our streets.''} According to Ehrenfest's assistant Casimir who was in the
    audience, Ehrenfest improvised something like this: \textit{``and you might also considerably
    reduce the expenditure for a beautiful, new, formal black suit''} (quoted in \cite{Enz},
    p.258).

     These remarks indicate the role of the exclusion principle for the stability of matter in
    bulk. A lot of insight and results on this central issue, both for ordinary matter
    (like stones) and self-gravitating bodies, have been obtained in more recent times, beginning with
    the work of Dyson and Lenard in 1967 \cite{DL}. For further information, I highly recommend the review articles in Lieb's Selecta \cite{L1}. (For a brief description, see \cite{NS1}.)

\section{Pauli's discovery of the relation between matrix mechanics and wave mechanics (letter to P. Jordan)}

On April 12, 1926 Pauli wrote a very remarkable letter to P. Jordan (\cite{Mey}, Vol. I, letter 131), just after the first communication of Schr\"{o}dinger had appeared.


B. L. van der Waerden devoted his talk at the ``Dirac conference'' in Trieste in 1972 \cite{Mer} almost entirely to this letter. In this Pauli established the connection between wave and matrix mechanics in a \textit{``logically irreproachable way, independent of Schr\"{o}dinger. He never published the contents of this letter, but signed a carbon copy (which is quite unusual) and he kept the letter in a plastic cover until his death''} (van der Waerden's words).

I would like to go through this letter, which is also remarkable in other respects. At the same time it gives an impression of the enormous influence Pauli had through his extensive correspondence. Pauli's letters are an integral part of his work and thinking. It is also a wonderful experience to read at least some of them. The letter begins with

\begin{quote}

\textit{``Dear Jordan,\\
Many thanks for your last letter and for looking through the proof sheets. Today I want to write neither about my Handbuch-Article nor about multiple quanta; I will rather tell you the results of some considerations of mine connected with Schr\"{o}dinger's paper `Quantisierung als Eigenwertproblem' which just appeared in the \textit{Annalen der Physik}. I feel that this paper is to be counted among the most important recent publications. Please read it carefully and with devotion.\\
Of course I have at once asked myself how his results are connected with those of the G\"{o}ttingen Mechanics. I think I have now completely clarified this connection. I have found that the energy values from Schr\"{o}dinger's approach are always the same as those of the G\"{o}ttingen Mechanics, and that from Schr\"{o}dinger's functions $\psi$, which describe the eigenvibrations, one can in a quite simple and general way construct matrices satisfying the equations of the G\"{o}ttingen Mechanics. Thus at the same time a rather deep connection between the G\"{o}ttingen Mechanics and the Einstein-de Boglie Radiation Field is established.\\
To make this connection as clear as possible, I shall first expose Schr\"{o}dinger's approach, styled a little differently.``
}
\end{quote}

Pauli does not start with Schr\"{o}dinger's stationary equation of his `First Communication', whose justification I find, by the way, rather obscure\footnote{A profound justification, based on the mechanical-optical analogy, was given in the Second Communication.}. Pauli first derived what we now call the Klein-Gordon equation\footnote{From Schr\"{o}dinger's research notes we know that he studied this equation before he had the Schr\"{o}dinger equation, but abandoned it because it gave the wrong fine structure for hydrogen.}. He starts from the relativistically invariant Einstein-de Broglie relations $\mathbf{p}=\hbar\mathbf{k},~E=\hbar\omega$, and inserts these into the relativistic mechanical equation
\[ E-V=E_{kin}=\sqrt{c^2\mathbf{p}^2+(mc^2)^2}~~\Rightarrow \mathbf{p}^2=\hbar^2\mathbf{k}^2=\frac{1}{c^2}\bigl[(E-V)^2-(mc^2)^2\bigr].\]
If this is inserted into the stationary wave equation
\[(\triangle + k^2)\psi=0 \]
one obtains the stationary Klein-Gordon equation
\[ \Bigl[\triangle+\bigl(\frac{E-V}{\hbar c}\bigr)^2-\bigl(\frac{mc}{\hbar}\bigr)^2\Bigr]\psi=0.  \]
(Actually, Pauli first arrives at a time-dependent equation, which is, however, different from the Klein-Gordon equation, except in the free case.) Then Pauli considers the non-relativistic limit and writes: \textit{``This equation is given in Schr\"{o}dinger's paper, and he also shows how it can be derived from a Variational Principle''.} After a remark about the analogy with the difference between Geometrical Optics and Wave Optics, he says:

\textit{``Next comes my own contribution, namely the connection with the G\"{o}ttin\-gen Mechanics.''} With the complete set of eigenfunctions of the Schr\"{o}dinger equation he associates to operators of the Schr\"{o}dinger theory matrices, and verifies that these \textit{``satisfy the equations of the G\"{o}ttingen Mechanics''.} Since we are all familiar with this, no further comments are necessary.

During the short time after the First Communication of Schr\"{o}dinger had appeared, Pauli did more:
\begin{quote}
\textit{``I have calculated the oscillator and rotator according to Schr\"{o}din\-ger. Further the H\"{o}nl-Kronig-formulae for the intensity of Zeeman components are easy consequences of the properties of the spherical harmonics. Perturbation theory can be carried over completely into the new theory, and the same thing holds for the transformation to principle axes, which in general is necessary if degenerations (multiple eigenvalues) are cancelled by external fields of force. At the moment I am occupying myself with the calculation of transition probabilities in hydrogen from the eigenfunctions calculated by Schr\"{o}dinger. For the Balmer lines finite rational expressions seem to come out. For the continuous spectrum the situation is more complicated: the exact mathematical formulation is not yet quite clear to me.''}
\end{quote}

On this Pauli was again a bit too late to submit a paper. The one by Schr\"{o}dinger was submitted on March 18, who wrote a month later to Sommerfeld: \textit{``Mit Pauli habe ich ein paar lange Briefe gewechselt. Er ist schon ein ph\"{a}nomenaler Kerl. Wie der wieder alles schnell heraussen gehabt hat! In einem Zehntel der Zeit, die ich dazu gebraucht hab.''}

The letter of Pauli to Jordan ends with: \textit{``Cordial greetings for you and the other people in G\"{o}ttingen (especially to Born, in case he is back from America; please show him this letter).''}

\paragraph{Supplementary remarks.} From a letter of Pauli to Schr\"{o}dinger late in 1926 it is clear that he independently discovered the gauge invariance. In this letter Pauli begins by saying that at first sight the relativistic wave equation does not only contain the field strengths, but also the absolute values of the 4-potential. However, he adds: \textit{``Thanks God this is only apparent''}, and he gives the formulae for what we call gauge invariance of the relativistic Kein-Gordon equation. Again, Pauli did not publish the content of this letter, because he learned from Schr\"{o}dinger's answer about a paper Schr\"{o}dinger had just submitted to the `Annalen' (two days before Pauli had written his letter). However, Schr\"{o}dinger says in his paper nothing about gauge invariance.

\section{On Pauli's invention of non-Abelian Kaluza-Klein Theory in 1953}

There are documents which show that Wolfgang Pauli constructed in 1953 the first consistent
generalization of the five-dimensional theory of Kaluza, Klein, Fock and others to a higher dimensional
internal space. Because he saw no way to give masses to the gauge bosons, he refrained from publishing
his results formally. This is still a largely unknown chapter of the early history of non-Abelian gauge
and Kaluza-Klein theories.

 Pauli described his detailed attempt of a non-Abelian generalization of Kaluza-Klein theories extensively in some letters to A. Pais, which have been published in Vol. IV, Part II of Pauli's collected letters \cite{kn:Pau1}, as well in two seminars in Z\"urich on November 16 and 23, 1953. The latter have later been  written up in Italian by Pauli's pupil P. Gulmanelli \cite{kn:Gul}. An English translation of these notes by P. Minkowski is now available on his home page. By specialization (independence of spinor fields on internal space) Pauli got all important formulae of Yang and Mills, as he later (Feb. 1954) pointed out in a letter to Yang \cite{kn:Pau6}, after a talk of Yang in Princeton. Pauli did not publish his study, because he was convinced that \textit{"one will always obtain vector mesons with rest mass zero"} (Pauli to Pais, 6 Dec., 1953).

\subsection{The Pauli letters to Pais}

At the Lorentz-Kammerlingh Onnes conference in Leiden (22-27 June 1953) A. Pais talked about an attempt of describing nuclear forces based on isospin symmetry and baryon number conservation. In this contribution he introduced fields, which do not only depend on the spacetime coordinates $x$, but also on the coordinates $\omega$ of an internal isospin space. The isospin group acted, however, globally, i.e., in a spacetime-independent manner.\\
During the discussion following the talk by Pais, Pauli said:
\begin{quote}
\textit{``...I would like to ask in this connection whether the transformation group with constant phases can be amplified in a way analogous to the gauge group for electromagnetic potentials in such a way that the meson-nucleon interaction is connected with the amplified group...''}
\end{quote}

Stimulated by this discussion, Pauli worked on the problem, and wrote on July 25, 1953 a long technical letter to Pais \cite{kn:Pau2}, with the motto: "Ad \textit{usum Delfini} only". This letter begins with a personal part in which Pauli says that \textit{"the whole note for you is of course written in order to drive you further into the real virgin-country"}. The note has the interesting title:\\
\, \\
\textit{``Written down July 22-25 1953, in order to see how it looks. Meson-Nucleon Interaction and Differential Geometry.''}\\
\, \\
In this manuscript, Pauli generalizes the original Kaluza-Klein theory to a six-dimensional space and arrives through dimensional reduction at the essentials of an $SU(2)$ gauge theory. The extra-dimensions form a two-sphere $S^2$ with space-time dependent metrics on which the $SU(2)$ operates in a space-time-dependent manner. Pauli emphasizes that this transformation group \textit{"seems to me therefore the \textsf{natural generalization of the gauge-group} in case of a two-dimensional spherical surface".} He then develops in 'local language' the geometry of what we now call a fibre bundle with a homogeneous space as typical fiber (in this case $SU(2)/U(1)$).\\
Since it is somewhat difficult to understand exactly what Pauli did, we give some details, using more familiar formulations and notations \cite{kn:Strau}.\\
Pauli considers the six-dimensional total space $M \times S^2$, where $S^2$ is the two-sphere on which $SO(3)$ acts in the canonical manner. He distinguishes among the diffeomorphisms (coordinate transformations) those which leave the space-time manifold $M$ pointwise fixed and induce space-time-dependent rotations on $S^2$:
\begin{equation}
(x,y)\rightarrow [x,R(x)\cdot y].
\end{equation}
Then Pauli postulates a metric on $M \times S^2$ that is supposed to satisfy three assumptions. These led him to what is now called the non-Abelian Kaluza-Klein ansatz: The metric $\hat{g}$ on the total space is constructed from a space-time metric $g$, the standard metric $\gamma$ on $S^2$, and a Lie-algebra-valued 1-form,
\begin{equation}
A=A^a T_a \, , \,  A^a=A^a_\mu dx^\mu,
\end{equation}
on $M$ ($T_a$, $a=1,2,3$, are the standard generators of the Lie algebra of $SO(3)$) as follows: If $K^i_a \partial / \partial y^i$ are the three Killing fields on $S^2$, then
\begin{equation}
\hat{g}=g-\gamma_{ij}[dy^i+K^i_a(y)A^a]\otimes[dy^j+K^j_a(y)A^a].
\end{equation}
In particular, the non-diagonal metric components are
\begin{equation}
\hat{g}_{\mu i}=A^a_\mu(x)\gamma_{ij}K^j_a.
\end{equation}
Pauli does not say that the coefficients of $A^a_\mu$ in Eq.\;(4) are the components of the three independent Killing fields. This is, however, his result, which he formulates in terms of homogeneous coordinates for $S^2$. He determines the transformation behavior of $A^a_\mu$ under the group (1) and finds in matrix notation what he calls \textit{"the generalization of the gauge group"}:
\begin{equation}
A_\mu \rightarrow R^{-1} A_\mu R+R^{-1}\partial_\mu R.
\end{equation}
With the help of $A_\mu$, he defines a covariant derivative, which is used to derive \textit{"field strengths"} by applying a generalized curl to $A_\mu$. This is exactly the field strength that was later introduced by Yang and Mills. To our knowledge, apart from Klein's 1938 paper, it appears here for the first time. Pauli says that \textit{"this is the \textsf{true} physical field, the analog of the \textsf{field strength}"} and he formulates what he considers to be his \textit{"main result"}:\\
\, \\
\textit{The vanishing of the field strength is necessary and sufficient for the $A^a_\mu (x)$ in the whole space to be transformable to zero.}\\
\, \\
It is somewhat astonishing that Pauli did not work out the Ricci scalar for $\hat{g}$ as for the Kaluza-Klein theory. One reason may be connected with his remark on the Kaluza-Klein theory in Note\;23 of his relativity article\cite{P6} concerning the five dimensional curvature scalar (p.\;230):
\begin{quote}
\textit{There is, however, no justification for the particular choice of the five-dimensional curvature scalar $P$ as integrand of the action integral, from the standpoint of the restricted group of the cylindrical metric (gauge group). The open problem of finding such a justification seems to point to an amplification of the transformation group.}
\end{quote}
In a second letter \cite{kn:Pau4}, Pauli also studies the dimensionally reduced Dirac equation and arrives at a mass operator that is closely related to the Dirac operator in internal space $(S^2,\gamma)$. The eigenvalues of the latter operator had been determined by him long before \cite{kn:Pau5}. Pauli concludes with the statement: \textit{"So this leads to some rather unphysical \textsf{shadow particles}"}.

Pauli's main concern was that the gauge bosons had to be massless, as in quantum electrodynamics. He emphasized this mass problem repeatedly, most explicitly in the second letter \cite{kn:Pau4} to Pais on December 6, 1953, after he had made some new calculations and had given the two seminar lectures in Zurich already mentioned. He adds to the Lagrangian what we now call the Yang-Mills term for the field strengths and says that \textit{"one will always obtain vector mesons with rest-mass zero (and the rest-mass if at all finite, will always remain zero by all interactions with nucleons permitting the gauge group)." To this Pauli adds: "One could try to find other meson fields"}, and he mentions, in particular, the scalar fields which appear in the dimensional reduction of the higher-dimensional metric. In view of the Higgs mechanism this is an interesting remark.

Pauli learned about the related work of Yang and Mills in late February, 1954, during a stay in Princeton, when Yang was invited by Oppenheimer to return to Princeton and give a seminar on his joint work with Mills. About this seminar Yang reports \cite{kn:Ya}: "Soon after my seminar began, when I had written down on the blackboard $(\partial_\mu -i \epsilon B_\mu )\Psi$, Pauli asked: \textit{What is the mass of this field $B_\mu$?}, I said we did not know. Then I resumed my presentation, but soon Pauli asked the same question again. I said something to the effect that that was a very complicated problem, we had worked on it and had come to no conclusion. I still remember his repartee: `That is no sufficient excuse.' I was so taken aback that I decided, after a few moments' hesitation to sit down. There was general embarrassment. Finally Oppenheimer said, `we should let Frank proceed.' Then I resumed and Pauli did not ask any more questions during the seminar.'' (For more on this encounter, see \cite{kn:Ya}.)

In a letter to Yang \cite{kn:Pau6} shortly after Yang's Princeton seminar, Pauli repeats: \textit{"But I was and still am disgusted and discouraged of the vector field corresponding to particles with zero rest-mass (I do not take your excuses for it with 'complications' seriously) and the difficulty with the group due to the distinction of the electromagnetic field remains."} Formally, Pauli had, however, all important equations, as he shows in detail, and he concludes the letter with the sentence: "On the other hand you see, that your equations can easily be generalized to include the $\omega$-space" (the internal space). As already mentioned, the technical details have been written up by Pauli's pupil P. Gulmanelli \cite{kn:Gul}and have recently been translated by P. Minkowski from Italian to English.

\begin{center}
  * \qquad * \qquad *
\end{center}
I hope that my scattered remarks have at least indicated that Wolfgang Pauli was a great man of
uncompromising scientific honesty, to whom his own words \cite{P13} on Einstein apply equally well:
{\it ``His life anticipating the future will forever remind us of the ideal -- under
threat in our time -- of spiritual, contemplative man, his thoughts calmly and
unswervingly bent on the great problems of the structure of the cosmos.''}


\newpage

\begin{thebibliography}{99}

\bibitem{FW}
M. Fierz and V. Weisskopf Eds.), \textit{Theoretical physics in the twentieth century: A memorial volume to Wolfgang Pauli.} New York: Interscience (1960).

\bibitem{Mey}
K. von Meyenn (Ed), \textit{Wolfgang Pauli: Scientific Correspondence with Bohr, Einstein, Heisenberg, a.O., Vol. I-IV}, volume 2,6,11,14,15,17,18 of \textit{Sources in the History of Mathematics and Physical Sciences}. Springer Verlag Heidelberg and New York, 1979-2005.

\bibitem{P1}
W. Pauli, Writings on Physics and Philosophy. Springer Verlag, Berlin, 1994. Edited by Ch. P. Enz and K. von Meyenn, translated by R. Schlapp.

\bibitem{Enz}
Ch. P. Enz, \textit{No time to be brief: a scientific biography of Wolfgang Pauli.} Oxford University Press, New York 2002.

\bibitem{NS2}
    N. Straumann, {On Wolfgang Pauli's most important contributions to physics}. Opening
    talk at the symposium: WOLFGANG PAULI AND MODERN PHYSICS, in honor of the 100th
    anniversary of Wolfgang Pauli's birthday, ETH (Zurich), Mai 4-6 2000 [arXiv: physics/0010003].

\bibitem{P2}
W. Pauli jr., Physikalische Zeitschrift \textbf{20}, 25 (1919).

\bibitem{P3}
W. Pauli jr., Physikalische Zeitschrift \textbf{20}, 457 (1919).

\bibitem{P4}
W. Pauli jr., Verhandlungen der Deutschen Physikalischen Gesellschaft \textbf{21}, 742 (1919).

\bibitem{P5}
W. Pauli, in: Encyklop\"{a}die der mathematischen Wissenschaften, Vol. V/19, pp. 539-775 (1921), Teubner, Leipzig, 1921; New German Edition: Edited and annotated by D. Giulini, including Pauli's own supplementary notes from 1956, Springer Verlag, Berlin, 2000.

\bibitem{P6}
W. Pauli, \textit{Theory of Relativity}, Dover Publications, Inc., New York, 1981.

\bibitem{P7}
W. Pauli, {\it Remarks on the History of the Exclusion Principle}, Science \textbf{103}, 213-215 (1946).

\bibitem{NS1}
N. Straumann, \textit{The Role of the Exclusion Principle for Atoms to Stars: A Historical Account}, International Review of Physics, Vol. 1, 184-196 (2007) [arXiv:quant-ph/0403199].


\bibitem{P8}
    W. Pauli, {\it \"Uber den Zusammenhang des Abschlusses der
    Elektronengruppen im Atom mit der Komplexstruktur der
    Spektren}, Z.~Phys.~{\bf 31}, 765--783 (1925).


\bibitem{P9}
    W. Pauli, \textit{On the connection between spin and statistics}, Physical Review, \textbf{58}, 716 (1940).

\bibitem{Ein}
A. Einstein, {\it The Collected Papers of Albert Einstein}, Vols.
1-10, Princeton University Press, 1987--. See also: [http:www.
einstein.caltech.edu/].


\bibitem{P10}
W. Pauli, {\it Pauli Lectures on Physics}; Ed. C.P. Enz. MIT Press
(1973); Vol.\;4, especially Sect.\;20.


\bibitem{ET}
C.P. Enz, and A. Thellung, Helv. Phys. Acta \textbf{33}, 839 (1960).


\bibitem{P11}
W. Pauli, {\it Die allgemeinen Prinzipien der Wellenmechanik}.
Handbuch der Physik, Vol. XXIV (1933). New edition by N. Straumann,
Springer-Verlag (1990); see Appendix III, p. 202.



\bibitem{Sch}
    E.L. Schucking, {\it Jordan, Pauli, Politics, Brecht, and a variable gravitational
    constant}, Physics Today, October 1999, p. 26-31.


\bibitem{P12}
    W. Pauli, {\it Zur Quantenmechanik des magnetischen
    Elektrons}, Z. Physik {\bf 43}, 601--623 (1927).

\bibitem{E}
    P. Ehrenfest, {\it Ansprache zur Verleihung der Lorentzmedaille
    an Professor Wolfgang Pauli am 31.~Oktober
    1931.} Versl. Akad. Amsterdam {\bf 40}, 121--126 (1931).

\bibitem{DL}
    F.J. Dyson and A. Lenard, {\it Stability of Matter, I and II}, J. Math. Phys.
    \textbf{8}, 423-434 (1967); ibid \textbf{9}, 698-711 (1968).

\bibitem{L1}
    E.H. Lieb, {\it The Stability of Matter: From Atoms to Stars}, Selecta of Elliott H.
    Lieb, Springer-Verlag, 1991.

\bibitem{Mer}
B. L. van der Waerden, in:
J. Mehra (Ed.) \textit{The Physicist's Conception of Nature}, Reidel, dortrecht and Boston, 1973, p.\;315.

\bibitem{kn:Pau1}
W. Pauli, {\em Wissenschaftlicher Briefwechsel}, Vol. IV, Part II, 1999, Springer-Verlag, edited by K. V. Meyenn.

\bibitem{kn:Gul}
P. Gulmanelli, {\em Su una Teoria dello Spin Isotropico}, Pubblicazioni della Sezione di Milano dell'istituto Nazionale di Fisica Nucleare, Casa Editrice Pleion, Milano (1954).

\bibitem{kn:Pau2}
Pauli to Pais, Letter[1614] in \cite{kn:Pau1}.

\bibitem{kn:Strau}
L. O'Raifeartaigh and N. Straumann, {\em Reviews of Modern Physics} {\bf 72},
1-23 (2000).



\bibitem{kn:Pau4}
Pauli to Pais, Letter[1682] in \cite{kn:Pau1}.

\bibitem{kn:Pau5}
W. Pauli,
{\em Helv. Phys. Acta} {\bf 12}, 147 (1939).

\bibitem{kn:Ya}
C. N. Yang,
{\em Selected papers 1945-1980 with Commentary} (Freeman, San Francisco, 1983), p. 525.

\bibitem{kn:Pau6}
Pauli to Yang, Letter[1727] in \cite{kn:Pau1}.

\bibitem{P13}
W. Pauli, Translation from the article \textit{Impressionen \"{u}ber Albert Einstein} in: \textit{Aufs\"{a}tze und Vortr\"{a}ge  \"{u}ber Physik und Erkenntnistheorie}, Vieweg, Braunschweig (1961), p.81. Reprinted as Physik und Erkenntnistheorie, Vieweg und Braunschweig (1984).



\end{thebibliography}
\end{document}